\documentclass[pra, reprint,superscriptaddress,longbibliography]{revtex4-2}
\usepackage{amsmath}
\usepackage{amsfonts}
\usepackage{graphicx}
 \usepackage{physics}
 \usepackage[usenames]{color}

 \def\be{\begin{equation}} \def\ee{\end{equation}}
\def\bea{\begin{eqnarray}} \def\eea{\end{eqnarray}}

\begin{document}

	\date{\today}
	\title{Fast scrambling without appealing to holographic duality}
	
	\author{Zehan Li}
	\affiliation{Department of Physics and Astronomy, University of Pittsburgh, Pittsburgh, PA 15260, USA}
	
	\author{Sayan Choudhury}
	\email{sayan.choudhury@pitt.edu}
	\affiliation{Department of Physics and Astronomy, University of Pittsburgh, Pittsburgh, PA 15260, USA}
	
	\author{W. Vincent Liu}
	\email{wvliu@pitt.edu}
	\affiliation{Department of Physics and Astronomy, University of Pittsburgh, Pittsburgh, PA 15260, USA}
	\affiliation{Wilczek Quantum Center, School of Physics and Astronomy and T. D. Lee Institute, Shanghai Jiao Tong University, Shanghai 200240, China}
        \affiliation{Shanghai Research Center for Quantum Sciences, Shanghai 201315, China}
        \affiliation{Shenzhen Institute for Quantum Science and Engineering and Department of Physics, Southern University of Science and Technology, Shenzhen 518055, China}
	
	\begin{abstract}
Motivated by the question of whether all fast scramblers are holographically dual to quantum gravity, we study the dynamics of a non-integrable spin chain model composed of two ingredients - a nearest neighbor Ising coupling, and an infinite range $XX$ interaction. Unlike other fast scrambling many-body systems, this model is not known to be dual to a black hole. We quantify the spreading of quantum information using an out-of time-ordered correlator (OTOC), and demonstrate that our model exhibits fast scrambling for a wide parameter regime. Simulation of its quench dynamics finds that the rapid decline of the OTOC is accompanied by a fast growth of the entanglement entropy, as well as a swift change in the magnetization. Finally, potential realizations of our model are proposed in current experimental setups. Our work establishes a promising route to create fast scramblers.
	  
	\end{abstract}

	\maketitle

\section{Introduction} 

The dynamics of thermalization in closed quantum systems has received immense attention in recent years \cite{rigol2008thermalization,polkovnikov2011colloquium,leviatan2017quantum,neill2016ergodic,mallayya2019prethermalization,white2018quantum,nahum2017quantum,parker2019universal,mori2018thermalization}. A central focus of these studies has been the ``scrambling" of quantum information \cite{swingle2016measuring,yao2016interferometric,khemani2018operator,kelecs2019scrambling,garttner2018relating,marino2019cavity,xu2019locality,xu2020accessing,buijsman2017nonergodicity,alavirad2019scrambling,chavez2019quantum,chang2019evolution}. Scrambling is the process by which locally encoded information gets spread over non-local many-body degrees of freedom during the time evolution of a complex quantum system. This paradigm has been used to address a diverse array of questions in areas ranging from quantum chaos to quantum gravity \cite{aleiner2016microscopic,hosur2016chaos,chowdhury2017onset,roberts2015diagnosing,chenu2019work,lashkari2013towards,sekino2008fast,shenker2014black,maldacena2016bound,xu2020doesscrambling,geng2020non,bhagat2020generalized,choudhury2020cosmological}. Several recent experiments in a variety of analog quantum simulator platforms have successfully probed quantum scrambling \cite{garttner2017measuring,landsman2019verified,joshi2020quantum, li2017measuring,wei2019emergent,khurana2019unambiguous, blok2020quantum, meier2019exploring}, thereby paving the path to answer fundamental questions about non-equilibrium quantum dynamics.\\

Black holes are the fastest scramblers known in nature. Motivated by advances in holography, researchers have studied quantum many-body systems that can exhibit fast scrambling.  Perhaps, the most celebrated example of this is the Sachdev-Ye-Kitaev (SYK) model \cite{sachdev1993gapless,kitaev2015simple,maldacena2016remarks,gu2017local,davison2017thermoelectric}. This model describes $N$ Majorana fermions interacting via random infinite range interactions. The SYK model can be exactly solved in the large $N$ limit, where it is conjectured to be holographically dual to the Jackiw-Teitelboim model of gravity in two dimensions \cite{mandal2017coadjoint,jian2018quantum,gaikwad2020holographic}, and it can scramble as fast as a black hole in the low temperature limit \cite{sachdev2015bekenstein}. However, the randomness in the long range interactions is not necessary to produce fast scrambling. As Bentsen {\it et al.} have demonstrated in a recent paper, a non-disordered spin model describing sparsely connected spin-1/2 particles, can also be a fast scrambler \cite{bentsen2019treelike}. Their proposal was motivated by the p-adic version of the anti-de Sitter/conformal field theory correspondence \cite{gubser2017p}. While their model is very elegant, its experimental realization can be very difficult when the system size becomes large. Thus, it is necessary to search for alternative approaches to realize fast scrambling without disordered interactions. Furthermore, all of these works raises a crucial question: are all fast scramblers holographically dual to quantum gravity? \\

\begin{figure}
		\includegraphics[scale=0.6]{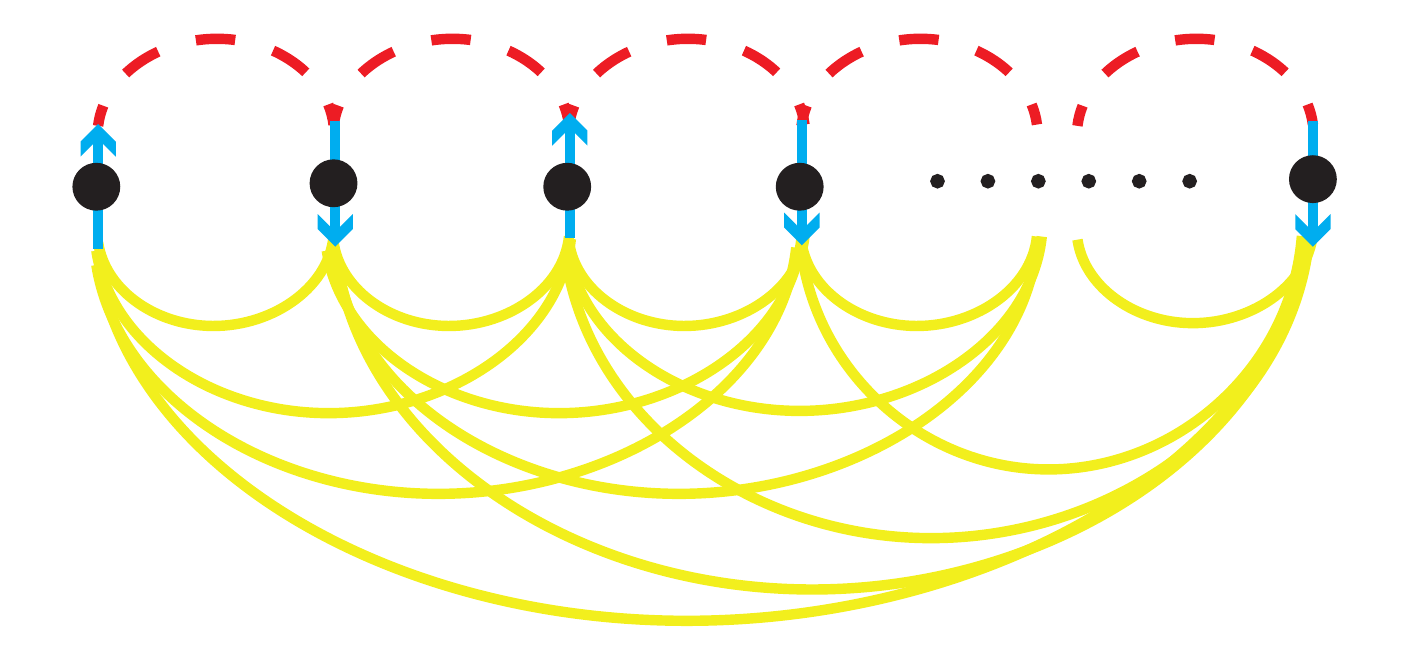}
		\caption{{\bf Schematic representation of the model:} The model in Eq.~(\ref{model}) is characterized by a nearest neighbor Ising coupling and an infinite range $XX$ coupling.}
		\label{fig_LevelStats}
\end{figure}

In this paper, we address this issue by proposing a fast scrambling many-body model, that is not inspired by holography. Our model essentially comprises two ingredients - a short range Ising interaction, and an infinite range $XX$ interaction. Both of these features are crucial since short range interacting systems can not be fast scramblers \cite{lieb1972finite}, while uniform infinite range interactions can not induce quantum chaos \cite{bentsen2019integrable}. Although our model is integrable in certain limits, we show that there is a large parameter regime, where the system exhibits fast scrambling. Furthermore, such a vanilla model may be easier to realize experimentally, even for large system sizes. Our results suggest that an appropriate combination of short and long range interactions can lead to fast scrambling. \\

We study information scrambling by studying the dynamics of an out-of-time ordered correlator (OTOC). In quantum chaotic systems, the growth of the OTOC at early times is exponential ($\sim e^{\lambda t}$), where $\lambda$ is bounded ($\lambda \le 2 \pi k_B T/\hbar$)\cite{maldacena2016bound}. Moreover, the fast scrambling conjecture states that the time it takes for local information be thoroughly scrambled in a $N$ body quantum system obeys a lower bound ($t \ge \log(N)/\lambda$) \cite{sekino2008fast}. We identify a large parameter regime where our model behaves as a fast scrambler. We also find that in this regime, our model is non-integrable, and the entanglement entropy grows very fast. \\

\section{Model} 

We study a one dimensional spin chain with $N$ sites described by the following Hamiltonian:
	\begin{equation}\label{model}
	H =\sum_{i=1}^{N} \left(\sigma_i^z \sigma^z_{i+1} + J  \sum_{j>i} (\sigma_i^+ \sigma^-_{j}+\sigma_i^- \sigma^+_{j}) \right),
	\end{equation}
where $\sigma_i^{\pm} = \frac{1}{\sqrt{2}}(\sigma_i^x \pm i \sigma_i^y)$ and $\sigma_i^{\gamma}$ is the standard Pauli matrix at lattice site $i$. A schematic of this model is shown in Fig.~\ref{fig_LevelStats}. In accordance with realistic experimental realizations of the all-to-all interaction, we do not rescale $J$ by $1/N$. We note that this spin chain can not exhibit fast scrambling, when $J$ is rescaled by $1/N$ (see Appendix A). Two other recent studies on related spin models have reached a similar conclusion \cite{belyansky2020minimal,yin2020bound}. When $J \rightarrow \infty$, this model reduces to a form of the Lipkin-Meshkov-Glick model, and it is mean field solvable in the thermodynamic limit \cite{lipkin1965validity,ribeiro2007thermodynamical}. On the other hand, when $J \rightarrow 0$, the model is the exactly solvable Ising model \cite{suzuki1967one}. Intriguingly, between these two extreme limits, this model can exhibit non-integrability - an essential criterion for fast scrambling (see Appendix A). In the remainder of the paper, we focus on the $J \sim O(1)$ regime, where the spin chain is characterized by Wigner-Dyson level statistics, and the system exhibits chaotic dynamics.\\

 \begin{figure}[h]
		\includegraphics[scale=0.37]{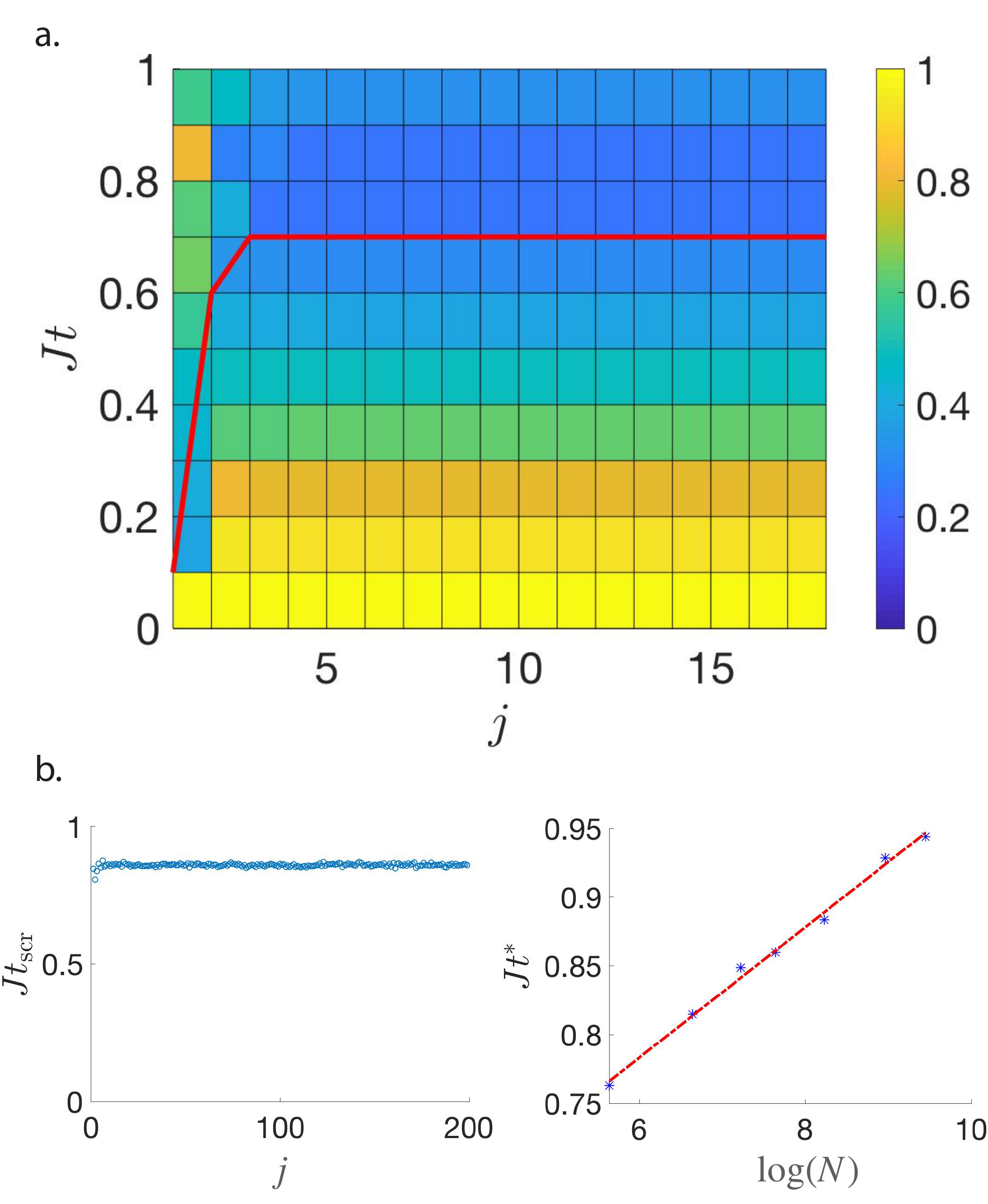}
		\caption{{\bf Scrambling of the infinite temperature state :} a. Time evolution of the OTOC, $F(j,t)$ (defined in Eq.~(\ref{otoceqn2})) for a $18-$site chain when $J=1$. The red line represents the time at which the OTOC reaches its minimum value. The OTOC spreads super-ballistically in this parameter regime. This is a salient characteristic of a fast scrambler. b. Semiclassical numerics for the dynamics of the spin chain when $J=1$. The left panel shows the time, $t_{\rm scr}$ at which the sensitivity $C_{\text cl} (j,t)$ (defined in Eq.~(\ref{eq_clsOTOC})), reaches $1$ on site $j$, when the chain length, $N=200$. We conclude that this system exhibits super-ballistic spreading, since $t_{\rm scr}$ is (almost) constant for $j \gg 1$. The right panel shows the system size dependence of the scrambling time $t^{*}$, at which $C_{\text cl} (j,t)$ reaches $1$ on all sites. We find that, $t^{*} \propto \log(N)$ - a characteristic signature of fast scrambling.}
		\label{fig_inftemp}
\end{figure}

While cousins of our model have been studied extensively \cite{santos2016cooperative,celardo2016shielding,lerose2019impact,lerose2018chaotic,zhu2019dicke,sundar2016lattice,chen2020persistent}, to the best of our knowledge, neither the equilibrium phase diagram, nor the non-equilibrium dynamics of this precise model has been studied before. Consequently, we discover a trove of rich non-equilibrium physics that arises from the interplay of nearest neighbor and infinite range interactions. We use exact diagonalization to study this model with open boundary conditions. The total $z$-magnetization ($M_z = \sum_{i=1}^N \sigma_i^z$) is conserved during the time evolution of this system, and we examine the $M_z = 0$ sector in this work. \\

\section{Out-of-time-order correlations}

Information scrambling is typically diagnosed by analyzing the dynamics of out-of-time ordered correlators (OTOCs). OTOCs capture the spreading of quantum information in a system by measuring operator growth. In particular, for two unitary and Hermitian operators A and B, the operator growth can be quantified by examining the expectation value of a squared commutator:
\begin{equation} \label{otoceqn}
	C(t)=\langle [A(t),B]^2\rangle = 2- 2 {\text{Re}} [\langle  A(t)B A(t)B\rangle],
\end{equation}
where $A(t) = e^{itH} A e^{- itH}$, and the OTOC is $\langle  A(t)B A(t)B\rangle$. For our model, we take $A=\sigma^z_1$, $B=\sigma^z_j$, and we compute the following OTOC:
\begin{equation} \label{otoceqn2}
	F(j,t)=\frac{1}{2}\left(1+{\text{Re}} [\langle \sigma^z_1(t)(t)\sigma^z_{j}\sigma^z_1(t)\sigma^z_{j}\rangle]\right).
\end{equation}

We note that $F$ is a bounded function ($0 \le F \le 1$). In quantum chaotic systems, $F(t)$ decays exponentially at early times i.e. $F(t)=1- \epsilon e^{\lambda_L t}$, where  $\lambda_L$ is analogous to the Lyapunov exponent. A salient characteristic of fast scramblers is that $F(j,t)$ spreads super-ballistically and starts deviating significantly from 1 on all sites, at a time $t^{*} \propto \log(N)$. In order to compare the properties of our model to other fast scramblers that have previously been studied in the literature \cite{hosur2016chaos,bentsen2019treelike,fu2016numerical,bentsen2019fast}, we compute $F(j,t)$ for the infinite temperature state. Our results for a $18$ site chain is shown in Fig.~\ref{fig_inftemp}a. It is clear from these figures that super-ballistic spreading of the OTOC occurs in this system, when $J\sim 1$. As detailed in Appendix B, we can obtain an analytical understanding of this behavior using a short-time expansion. \\

	\begin{figure}
		\includegraphics[scale=0.57]{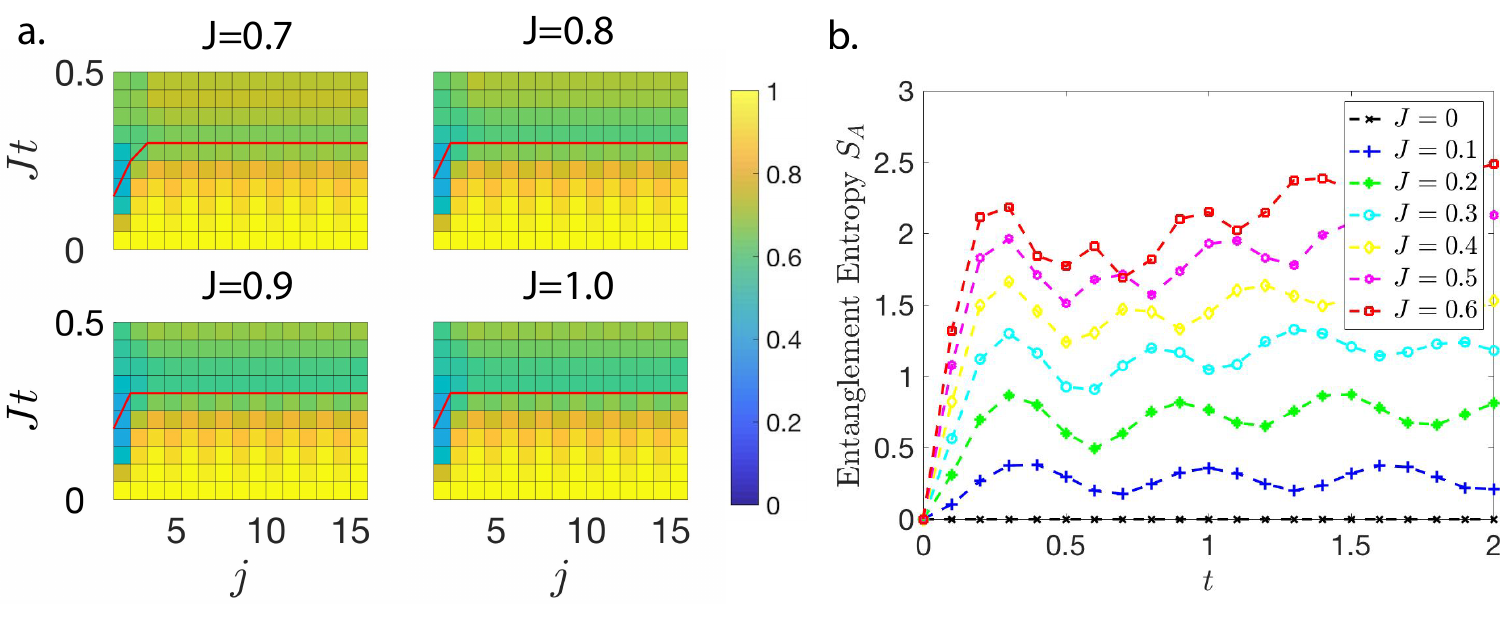}
		\caption{{\bf Quench dynamics for the classical N\'eel state: } a. The OTOC for different long range interactions, $J$. The red line is the time at which the OTOC reaches its minimal value. Similar to the infinite temperature case, the OTOC spreads super-ballistically. b. The dynamics of the half-chain entanglement entropy, for different long range interactions, $J$. This model is identical to the Ising model when $J=0$, and the entropy does not grow. As we increase $J$, the entropy grows faster and saturates to higher values. This result is consistent with the behavior of the OTOC.}
		\label{fig_FM}
	\end{figure}
	
While all the numerical results that we have discussed so far are exact, our study has been limited to small system sizes. However, in order to convincingly establish that out model is indeed a fast scrambler, we have to determine the dependence of $t^{*}$ on $N$ for $N \gg 1$. To overcome this limitation, we study the spin-$S$ version of our model, in the $S \rightarrow \infty$ limit, where it can be analyzed semi-classically. Following Ref.~\cite{cotler2018out}, we compute the averaged sensitivity:
    \begin{equation}\label{eq_clsOTOC}
        C_{\rm cl}(j,t)=\frac{1}{4 S^2}\langle \left(\frac{dS_j^z(t)}{d\phi}\right)^2\rangle,
    \end{equation}
    \noindent
where $\phi$ is a small initial rotation of spin $1$ about the $z$-axis, and the factor of $1/4$ has been introduced to establish correspondence with $F(j,t)$. This quantity can capture the sensitivity to initial conditions in classical systems, and it can be derived from Eq.~(\ref{otoceqn}) by substituting the commutator with appropriate  Poisson brackets. In order to compute the infinite temperature OTOC, we evaluate Eq.~(\ref{eq_clsOTOC}) for an ensemble in which each spin is initially aligned in a random direction. We characterize the scrambling rate in this semi-classical limit by computing the $j-$dependence of the time, $t_{\rm scr}$ at which $C_{\rm cl}(j,t_{\rm scr})$ becomes significant $(\sim 1)$. As shown in Fig.~\ref{fig_inftemp}b, we find that $t_{\rm scr}$ is (almost) constant for $j \gg 1$, thereby implying that this chain exhibits super-ballistic spreading. Furthermore, we systematically analyze the system size dependence of the scrambling time, and find that $C_{\text cl} (t^{*})$ becomes significant $(\sim 1)$ on all sites at time $t^{*} \propto \log(N)$. These calculations confirm that our model can exhibit fast scrambling.\\

\section{Quench Dynamics}  

While the infinite temperature OTOC dynamics provides compelling evidence for fast scrambling in our model, preparing this state in an experiment can be challenging. To alleviate this concern, we also study quench dynamics of this system for different initial states in the $M_z=0$ sector. In particular, we examine the OTOCs for unentangled product states of the form $\vert z_1;z_2;z_3\ldots z_L \rangle$, where $z_i$ is a spin polarized along the z-direction at site $i$ ($\uparrow$ or $\downarrow$). Motivated by experiments on cold atoms, we study the time evolution of the system, when it is initially prepared in the classical N\'{e}el initial state ($\vert \uparrow \downarrow \uparrow \downarrow \ldots \uparrow \downarrow \uparrow \downarrow \rangle$). As shown in Fig.~\ref{fig_FM}, we find that signatures of fast scrambling can be seen in the quench dynamics. We observe qualitatively similar behavior for other initial states (see Appendix C). We note that the infinite temperature results imply that fast scrambling can be observed for any typical initial state \cite{yin2020bound}.\\

A complementary approach to study information propagation in a quantum many-body system is to examine the growth of the half-chain entanglement entropy, $S_A = {\text Tr}[\rho_L \log(\rho_L)]$, where $\rho_L = {\text Tr_R} (\vert \psi \rangle \langle \psi \vert)$ is the reduced density matrix obtained by tracing over the degrees of freedom of one half of the chain. As shown in Fig.~\ref{fig_FM}b, we find that $S_A$ grows faster and saturates to higher values as $J$ increases. These results agree with our previous observation that the system exhibits faster scrambling when the strength of the infinite range interaction increases, as long as the system remains non-integrable.\\

\begin{figure}
		\includegraphics[scale=0.5]{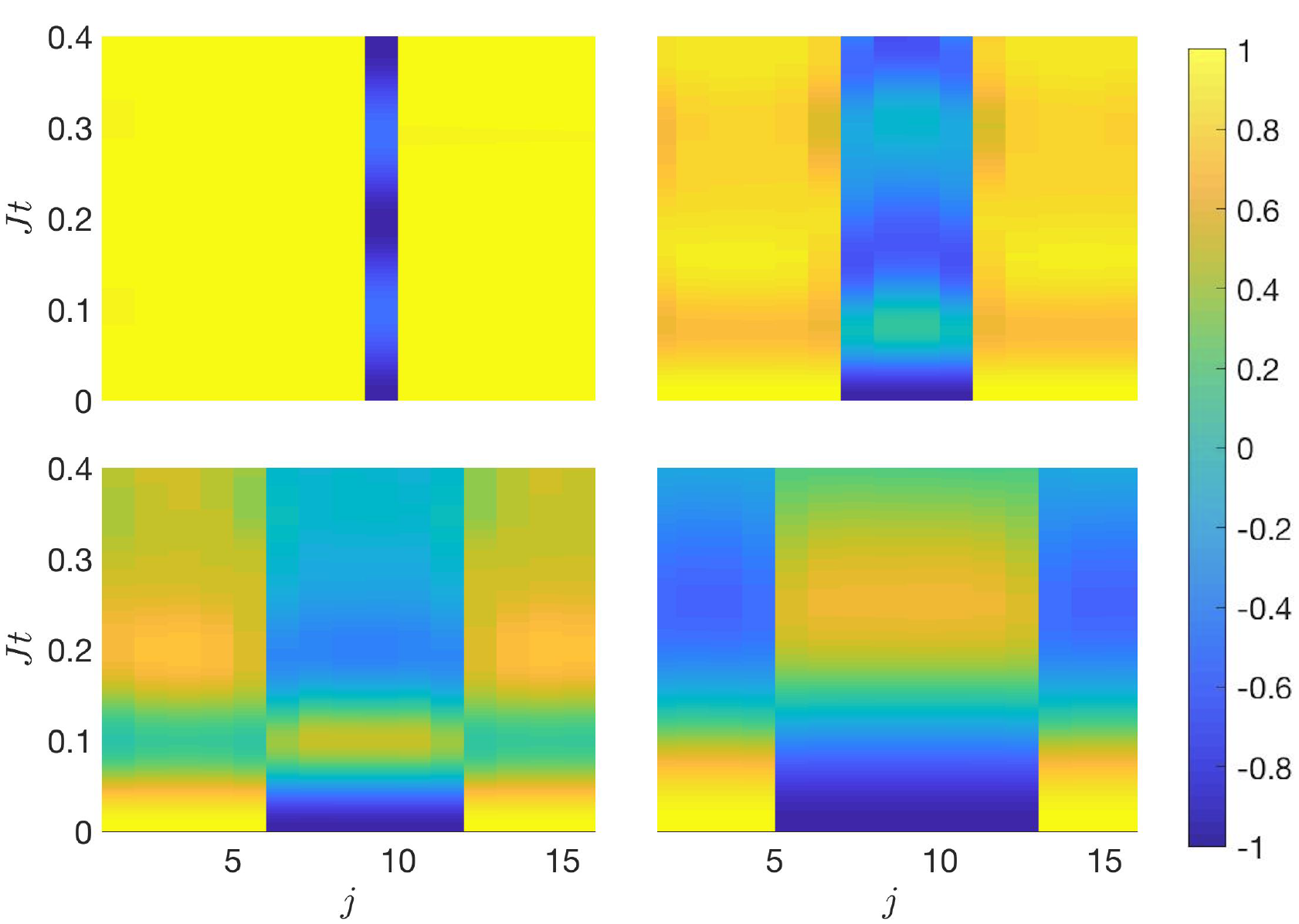}
		\caption{{\bf Quench dynamics for different magnetization sectors: }  Density plot for the local $z$-magnetization for initial states with different total magnetizations, when $J=1$. The initial state has $M$ spin downs clustered at the center of the chain, and  $N-M$ spin-ups present symmetrically around this cluster. We find that there is a crossover from slow to fast scrambling, as the spin imbalance decreases.}
		\label{fig_spin_imbalance}
	\end{figure}

Finally, we also explore the quench dynamics, when the total magnetization $M_z$ is finite. In particular, we compute the local magnetization, since this order parameter is accessible to experimental measurements. Furthermore, this quantity can be related to the recently proposed fidelity out-of time-ordered correlators, and can hence be used to quantify scrambling \cite{lewis2019unifying}. We note that in the one-magnon sector, i.e. when there is one spin-up (spin-down), and  $N-1$ spin-down (spin-up) in the initial state, then the system is integrable, and the system exhibits localized dynamics. This localization can be traced to the presence of localized eigenstates of the form: $\vert \psi \rangle = \vert \phi_i \rangle + \frac{1}{N-1} \sum_{j \ne i}  \vert \phi_j \rangle$ , where $\vert \phi_i \rangle = \vert \uparrow \uparrow\ldots \uparrow_{i-1} \downarrow_{i} \uparrow_{i+1}\ldots \uparrow \uparrow\rangle$ \cite{caruso2009highly}. While this model is integrable in the one-magnon sector, it can be non-integrable for a large parameter regime in the $N/2$ magnon sector. We carefully study the crossover from slow to fast scrambling as $|M_z|$ decreases. As shown in Fig.~\ref{fig_spin_imbalance}, we find that the fastest scrambling occurs when $M_z=0$. \\

\section{Experimental Realizations} 

The most natural way to realize our model is to couple an Ising spin chain to a single mode cavity \cite{ritsch2013cold}. By adiabatically eliminating the cavity degrees of freedom in the dispersive limit, we can obtain an effective infinite range coupling of the form described in Eq.~(\ref{model}) \cite{minavr2017effective,norcia2018cavity,davis2020protecting}. A promising scheme to realize our spin model using Rydberg dressed atoms in an optical cavity has been proposed in Ref.~\cite{gelhausen2016quantum}. As detailed in Appendix D, implementing this scheme is within the reach of current experiments. Another feasible route is to place a trapped ion crystal in the cavity \cite{gammelmark2011phase,luo2016dynamic}. Alternatively, it is possible to engineer this model by employing photon-mediated interaction between spins trapped in a photonic crystal waveguide \cite{hung2016quantum}, or by performing digital-analog simulations with trapped ions \cite{arrazola2016digital}. Several experimental protocols have been proposed to measure OTOCs in the experimental platforms described above. The infinite temperature OTOC can be determined by examining statistical correlations between measurements on randomized initial states \cite{vermersch2019probing}. Furthermore, some recent investigations have shown that it is possible to probe the scrambling dynamics after a quantum quench by measuring two point correlation functions \cite{lantagne2020diagnosing,blocher2020measuring}. Alternatively, interferometric techniques can also be used to measure OTOCs in different experimental platforms.\\

\section{Comparison with other fast scramblers} 

Before we conclude this paper, it is instructive to compare our model to other fast scramblers studied in the literature. As illustrated in Fig.~2, the scrambling time in our model is finite at infinite temperature. This feature is shared by several other fast scramblers, including most noticeably the SYK model, which describes $N$ Majorana fermions interacting via disordered global interactions drawn from a gaussian distribution of width $\mathcal{J}/N^{3/2}$. The scrambling time in this model shows a logarithmic dependence on the system size: $t^{*}\sim \log(N)/\mathcal{J}$ \cite{kobrin2020many,lucas2020non}. It is worth noting, there are some quantum chaotic systems, where the scrambling time is much faster ($\sim 0$) and $N$-independent \cite{bentsen2019fast}. A particularly striking example of this is a model of $N$ spins interacting with a central spin on a star like graph, where $t^{*}\sim 0$, even at finite temperatures \cite{lucas2019quantum}; this is the fastest scrambler found till date.\\

We note that after our preprint was posted on the arXiv, there appeared two other papers exploring fast scrambling in related spin models.  Belyansky {\it et al.} have demonstrated super-ballistic spreading of OTOCs in a spin model similar to ours (with the infinite range interaction is given by $J/\sqrt{N}$). Furthermore, they have argued that the infinite temperature Lyapunov exponent is finite in these systems and $t^{*}\sim 3/2 \log(N)/J^2$ \cite{belyansky2020minimal}. In a similar vein, Yin and Lucas have studied a family of non-integrable spin chains with two ingredients: (1) a global interaction rescaled by $1/N^{\delta}$ and (2) a time dependent magnetic field that ensures locally chaotic dynamics. They have derived a lower bound on the scrambling time in these models: $t^{*}>N^{\delta-1/2}$; fast scrambling occurs only when $\delta \le 1/2$ \cite{yin2020bound}. Thus, akin to our model, fast scrambling and extensivity of the total energy (which requires $\delta > 1$) can not occur simultaneously in these systems. Intriguingly, the scrambling rate can increase dramatically in these models (and even become $N$-independent), when the infinite range couplings are strongly time-dependent.\\

\section{Summary and Outlook} 
The paradigm of fast scrambling is of fundamental importance in understanding the dynamics of highly chaotic quantum systems. Observing fast scrambling is widely considered to be an important milestone towards exploring aspects of quantum gravity in the laboratory \cite{brown2019quantum,yang2020simulating,plugge2020revival}. Furthermore, fast scramblers can be harnessed for performing quantum information processing tasks, and is thus of great practical use \cite{brown2016holographic,brown2016complexity}. In this letter, we have demonstrated a novel route for creating a fast scrambler in an experimentally realizable spin model. Our proposal exploits the interplay of short and long range interactions to make the system highly chaotic.\\

By studying the infinite temperature OTOC of the system, using both exact diagonalization, and a semi-classical approximation technique, we have first demonstrated that the system exhibits fast scrambling. Next, we have examined the quench dynamics of the system, when it is initially prepared in the classical N\'{e}el state, and found that the OTOC and the half chain entanglement entropy grows very fast. Similar results are found for other non-entangled initial states when the total magnetization, $M_z$ is $0$. We have systematically explored how the scrambling rate depends on the total magnetization, and found that the system exhibits a crossover from slow to fast scrambling as the total magnetization decreases from $|M_z|=N$ (i.e. the fully polarized state) to $M_z=0$. Finally, we have proposed possible experimental realizations of our model. Thus, our work presents a rare example of a many-body model where the fast scrambling is not induced by random long range interactions, and provides a possible solution to the critical outstanding challenge of observing fast scrambling experimentally. \\

An extremely important feature of our work is that unlike other proposals studied in the literature, our model is not motivated by holography. This leads us to conjecture that fast scrambling can arise in non-holographic quantum matter. A rigorous proof of this conjecture can lead to the discovery of precise probes for distinguishing holographic and non-holographic quantum models, thereby shedding light on some fundamental questions in non-equilibrium quantum dynamics. Future work can examine other models with both short and long range interactions, and determine general conditions under which quantum many-body systems can exhibit fast scrambling.\\

\section*{Acknowledgments}
This work is supported by the AFOSR Grant No. FA9550-16-1-0006, the MURI-ARO Grant No. W911NF17-1-0323, the Shanghai Municipal Science and Technology Major Project (Grant No. 2019SHZDZX01), and the University of Pittsburgh Center for Research Computing through the resources provided.

\appendix

\begin{figure*}
		\includegraphics[scale=0.5]{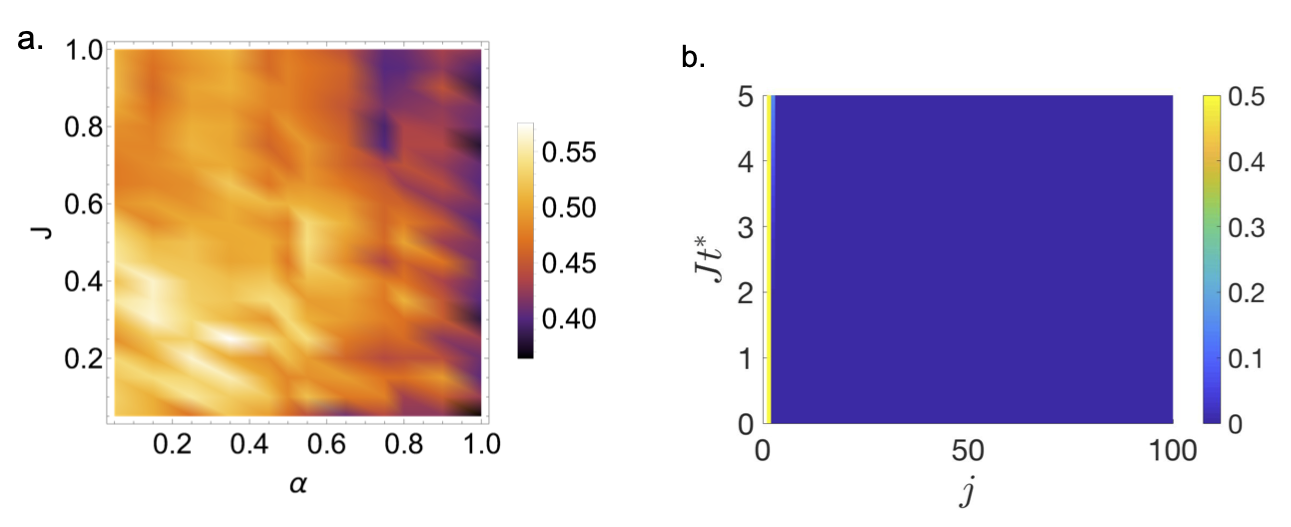}
		\caption{{\bf Level Statistics and scrambling of the infinite temperature state:}. a: The spectral statistics for our model (Eq.~\ref{modelsup}) when the total z-magnetization is 0, as characterized by the averaged ratio of adjacency gaps. We conclude that there is a large parameter regime, where $\langle r \rangle \sim 0.53$, and the system is non-integrable. We find that $\langle r \rangle \sim 0.39$, only when $\alpha \sim 1$, and the model is integrable. Fast scrambling is only expected when the system in non-integrable, and thus we focus on the $\alpha=0$ regime in the main text. b: The spread of the semiclassical sensitivity, $C_{\text cl} (j,t)$ for a $100$-site chain, when the infinite range interaction is $1/N$. In this case, the spin model does not exhibit fast scrambling. }
		\label{fig_ls}
\end{figure*}

\section{Level Statistics and Information Scrambling} 
An important diagnostic that is used to distinguish quantum chaotic systems from integrable systems is the energy level statistics \cite{santos2010onset,borgonovi2016quantum,d2016quantum}. Since fast scrambling can only occur when the system is non-integrable, we study the spectral statistics of our model  this section. To keep our discussion slightly more general, we study the following model:\\

\begin{eqnarray}\label{modelsup}
	H &=&\sum_{i=1}^{N}\sigma_i^z \sigma_{i+1}^z + \alpha (\sigma_i^+ \sigma^-_{i+1}+\sigma_i^- \sigma^+_{i+1}) \nonumber \\
	&+& J  \sum_{j>i} (\sigma_i^+ \sigma^-_{j}+\sigma_i^- \sigma^+_{j}) ,
	\end{eqnarray}
\\

We examine the level statistics of this model by sorting the energy eigenvalues $E_1<E_2<E_3<\ldots$, computing the adjacent energy gaps $\Delta E_n = E_{n+1}-E_n$, and then calculating the ratio of the adjacent energy gaps, $r_n = {\rm min} (\Delta E_m, \Delta E_{m+1})/{\rm max} (\Delta E_m, \Delta E_{m+1})$. Integrable systems are typically characterized by a Poisson distribution of $r_n$ i.e. $P(r) = 2/(1+r)^2$, with a mean value of $\langle r \rangle \approx 0.39$. In contrast, thermalizing systems are characterized by Wigner-Dyson distribution of $r_n$ i.e. $P(r) = (27/8)(r+r^2)/(1+r+r^2)^{5/2}$, with a mean value of $\langle r \rangle \approx 0.53$. Figure~\ref{fig_ls}a shows the energy level statistics for our model. We conclude that there is a wide parameter regime, where  $\langle r \rangle \sim 0.53$, and the system is thermalizing in nature; the model is integrable only when $\alpha \sim 1$. While, we have focused on $\alpha=0$ regime in the main text, we note that a small finite $\alpha$ does not alter the results qualitatively. \\

We now proceed to investigate the dynamics of the spin chain in the $\alpha=0$ regime when $J$ is rescaled by $1/N$. In this case, the system does not exhibit fast scrambling (see Fig.~\ref{fig_ls}b). While, we have presented the $N=100$ results here, we have verified that this result remains unchanged for other values of $J$ and $N$.  More generally, if $J$ is scaled by $1/N^{\alpha}$, then fast scrambling can occur only when $\alpha\le 1/2$ \cite{yin2020bound}.

\begin{figure*}
		\includegraphics[scale=0.75]{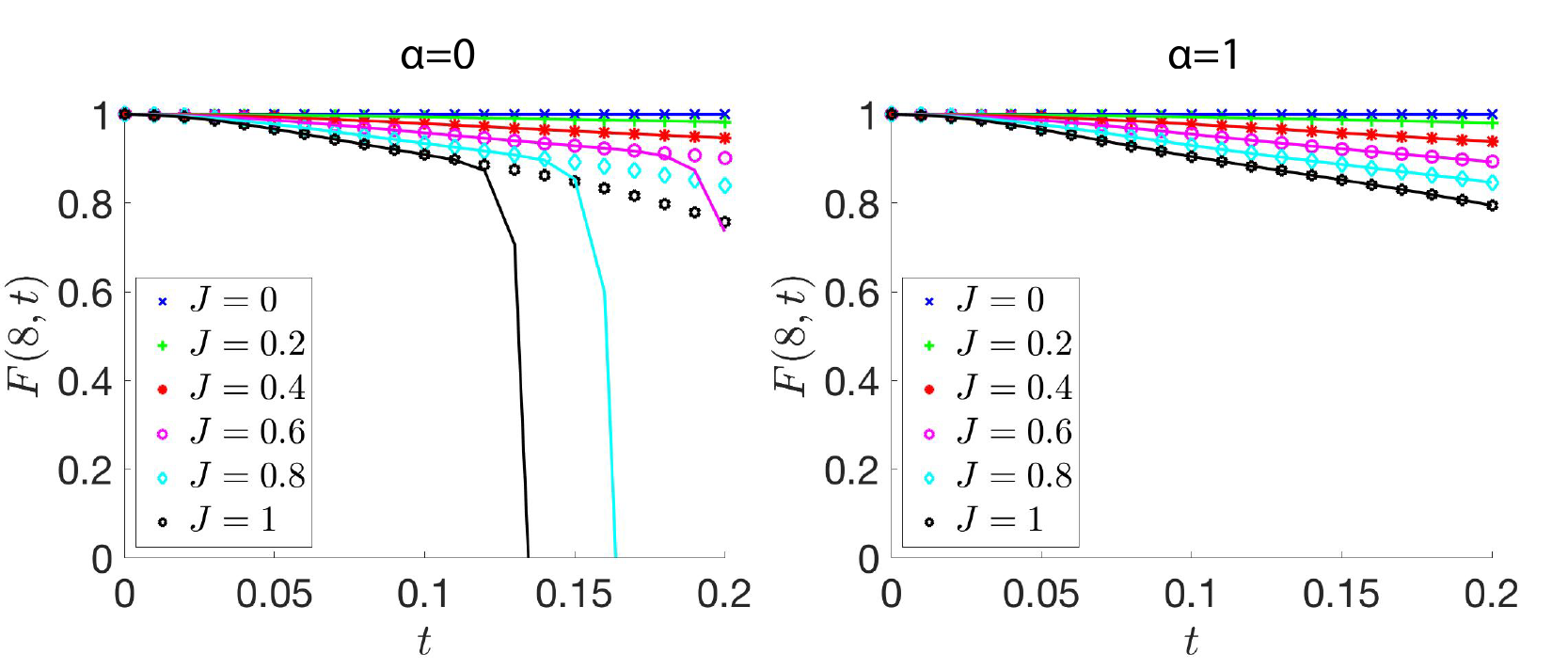}
		\caption{{\bf Comparison of an analytic short time expansion and exact diagonalization results for the OTOC, $F(8,t)$:} The circles represent numerical data from the exact diagonalizaton calculation, while the lines represent the analytical expression. Both approaches agree at short times, even though they differ at longer times in the fast scrambling regime.}
		\label{fig_TE}
\end{figure*}

\begin{figure*}
		\includegraphics[scale=0.65]{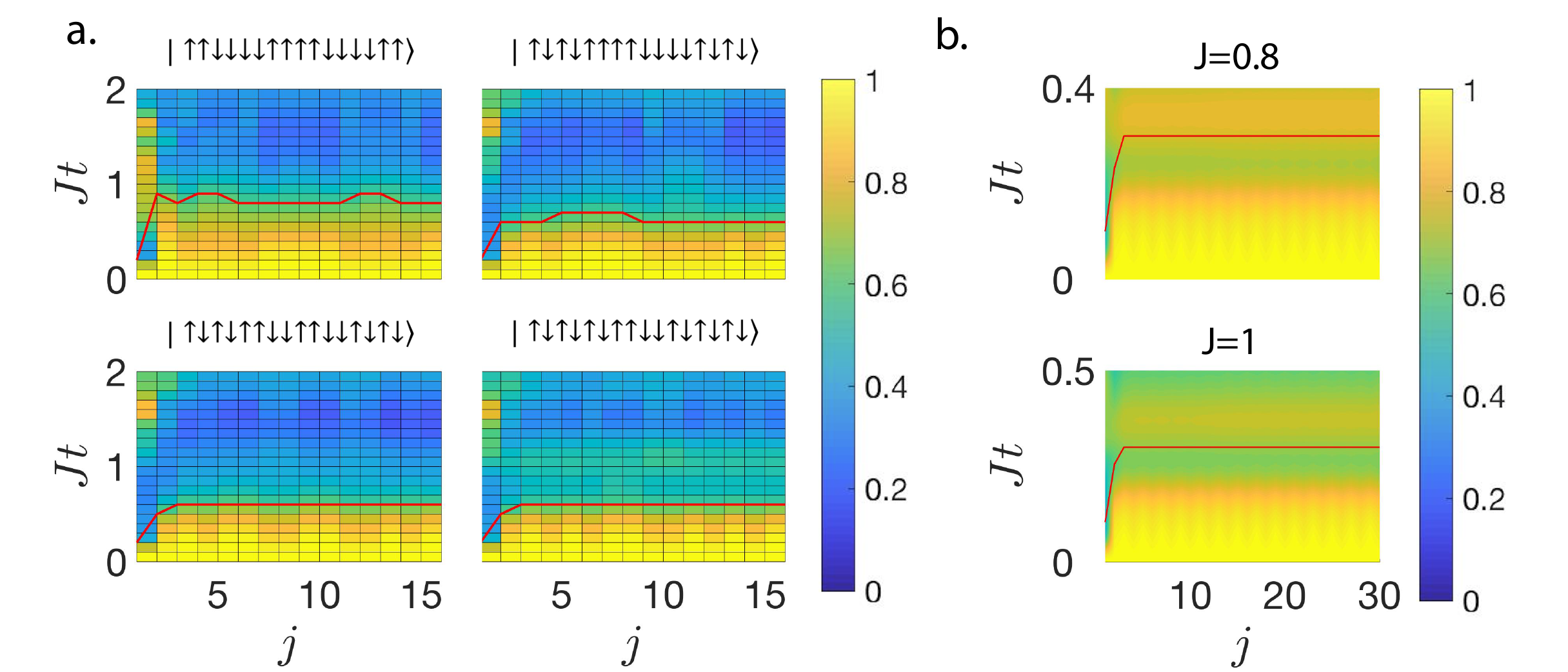}
		\caption{{\bf Quench Dynamics for $J=1$:} a: Exact results for the OTOC of a $16$-site chain initially prepared in various experimentally realizable product states. The initial states have been stated above each sub-figure. b: Matrix-product-state simulations for the quench dynamics of a $30$-site chain initialized in the classical N\'{e}el state. It is clear the the OTOC spreads super-ballistically in all of these cases.}
		\label{fig_quench}
\end{figure*}

\begin{figure}[h]
		\includegraphics[scale=0.25]{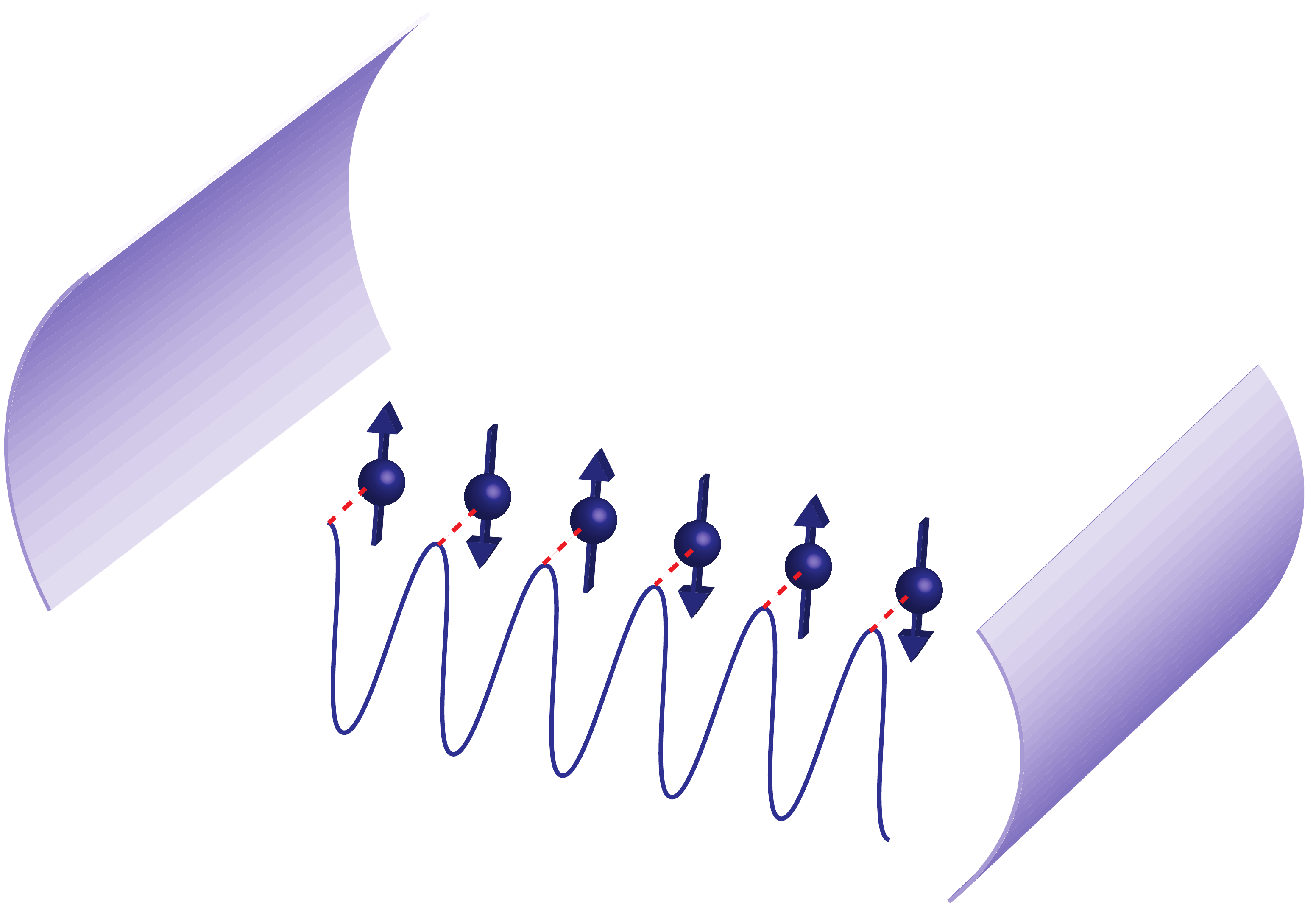}
		\caption{{\bf Schematic of the experimental realization of the spin model:} The fast scrambling model that we have studied can be realized when a one dimensional spin chain is collectively coupled to an optical cavity.}
		\label{fig_exset}
\end{figure}

\begin{figure}[h]
		\includegraphics[scale=0.3]{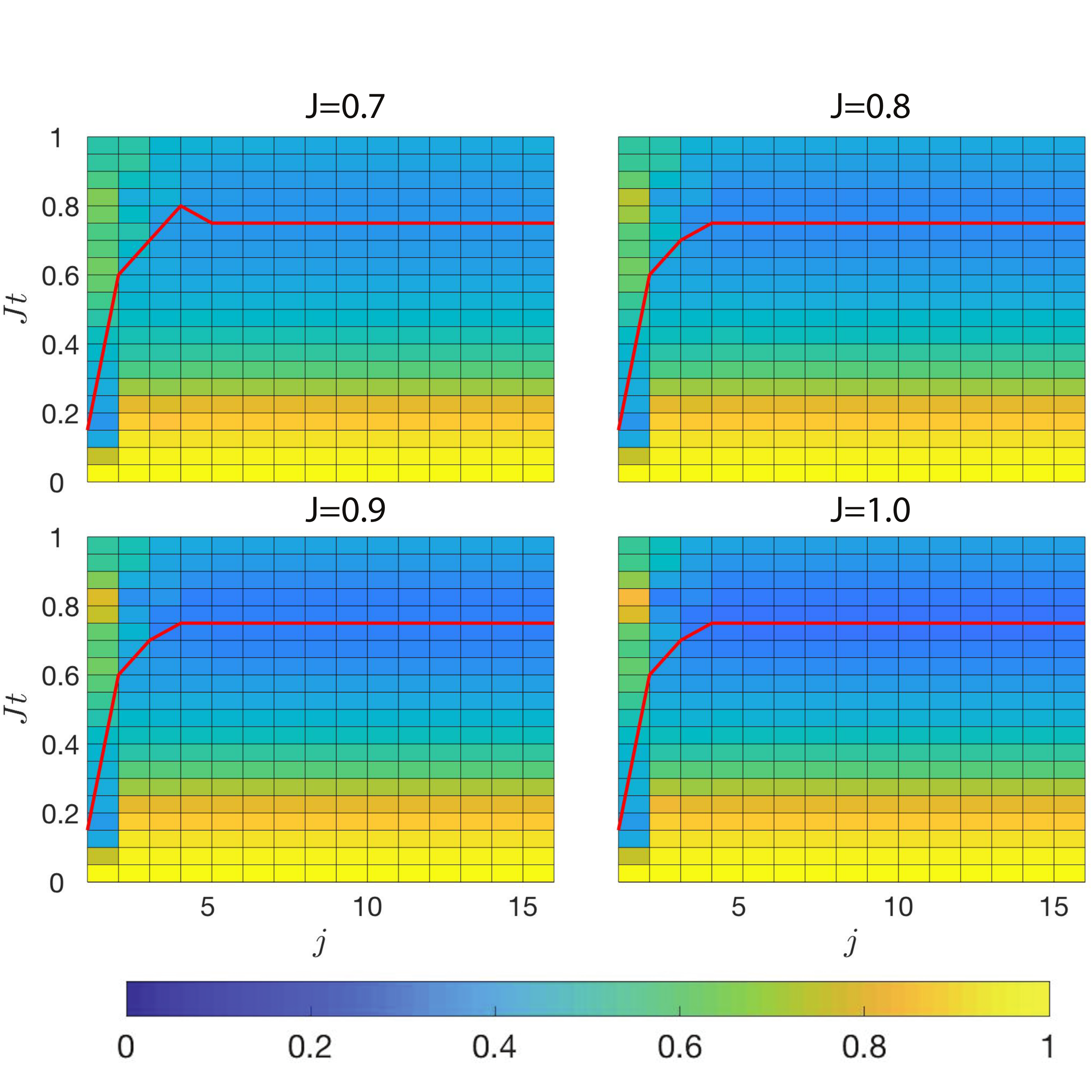}
		\caption{{\bf Fast scrambling of the infinite temperature state:} The OTOC for different values of the long range interactions, $J$, when $N=16$. The red line represents the time at which the OTOC reaches its minimum value. We find that when $J \sim O(1)$, the system exhibits super-ballistic spreading of the OTOC. The scrambling becomes faster with increasing $J$ in this regime.}
		\label{fig_last}
\end{figure}

\section{Short time expansion}
   
In the main text, we have computed the OTOC employing exact diagonalization. However, at early times, it is possible to obtain an analytical expression for the decay of the OTOC. To do this we expand the operator $\sigma_1^z(t)$ in the form
\begin{equation}\label{eq_TE}
	\sigma_1^z(t)=\sigma_1^z(0)-it[\sigma_1^z,H]-\frac{1}{2}t^2[[\sigma_1^z,H],H]+\ldots .
\end{equation}
	
Using Eq.~(\ref{eq_TE}), we can express the OTOC given in Eq.~(3) of the main text as a polynomial in $t$. Fig.~\ref{fig_TE} shows the comparison between the numerically calculated OTOC and analytical expression (upto $O(t^{30})$).\\

We find that for the non-integrable spin chain ($\alpha = 0$), there is reasonably good agreement between both approaches at short times. However, the analytical and numerical results diverge in the fast scrambling regime at longer times. The analytical expression is valid up to longer times, when the spin chain becomes integrable (i.e. $\alpha = 1$).\\

\section{Quench Dynamics}

We have already demonstrated that signatures of fast scrambling can be observed in the quench dynamics of the spin chain. In the main text, we had presented the results for the classical N\`{e}el initial state. However, we had concluded that this system is expected to exhibit similar dynamical behavior for any typical initial state \cite{yin2020bound}. In this section, we examine the dynamics of the model for some experimentally realizable initial states. By performing exact diagonalization on a $16$-site chain, we find that fast scrambling can indeed be exhibited by the spin chain for several initial product states (see Fig.~\ref{fig_quench}a). Furthermore, we employ matrix-product-state (MPS) techniques to access quantum dynamics for larger system sizes \cite{schollwock2011density}. As shown in Fig.~\ref{fig_quench}b, we find that a $30$-site chain initialized in the classical N\'{e}el state exhibits super-ballistic spreading. These results agree with the exact diagonalization calculations presented in the main text.

\section{Experimental Realization}
We have mentioned in the main text that the model in Eq.~(1) can be realized by coupling a spin chain to a single mode cavity \cite{gelhausen2016quantum}. In this section, we explicitly derive the effective spin Hamiltonian that arises when this scenario is realized. \\

The dynamics of an Ising chain interacting with a cavity can be described by the master equation:
     \begin{equation}
         \frac{d\hat{\rho}}{dt}=-i[\hat{H}_{SL},\hat{\rho}]+\mathcal{L}_{c}[\hat{\rho}],
     \end{equation}
where $\hat{\rho}$ is the density matrix of the system. The Hamiltonian describing the unitary evolution of the system is
\begin{equation}
         \hat{H}_{SL}=-\Delta_c \hat{a}^+ \hat{a}+ J_{\rm z} \sum_{i=1}^N  \hat{\sigma}_i^z 
         \hat{\sigma}_{i+1}^z+g \sum_{i=1}^N (\hat{a}^+\hat{\sigma}_i^- + \hat{a}\hat{\sigma}_i^+),
\end{equation}
where $\Delta_c$ is the effective cavity frequency, $g$ is the coupling between the spins and the cavity field, $J_{\rm z}$ is the Ising interaction strength, and the Lindblad term capturing the photon loss from the cavity at a rate $\kappa$ is given by:
\begin{equation}
         \mathcal{L}_c[\hat{\rho}]=\frac{\kappa}{2}(2\hat{a}\hat{\rho}\hat{a}^+ - \hat{a}^+\hat{a}\hat{\rho} - \hat{\rho}\hat{a}^+\hat{a}).
\end{equation}

We can eliminate the cavity mode adiabatically in the bad cavity limit ($\kappa \gg g$), and obtain a master equation for the reduced density matrix $\hat{\rho}_s$ of the spin chain,

\begin{equation}
         \frac{d\hat{\rho_s}}{dt}=-i[\hat{H}_{{\rm eff}},\hat{\rho}_s]+\mathcal{L}_{\Gamma}[\hat{\rho}_s],
\end{equation}

where the effective Hamiltonian is given by:
     \begin{equation}
         \hat{H}_{{\rm eff}}=\frac{4 g^2 \Delta_c}{4\Delta_c^2 + \kappa^2} \sum_{i,j}\hat{\sigma}_i^+\hat{\sigma}_j^- +  J_{\rm z} \sum_{i=1}^L \hat{\sigma}_i^z \hat{\sigma}_{i+1}^z,
     \end{equation}

and 
\begin{equation}
         \mathcal{L}_{\Gamma}[\hat{\rho}_s]= \frac{2 g^2 \kappa}{4\Delta_c^2 + \kappa^2}\sum_{i,j}(2\hat{\sigma}_i^-\hat{\rho}_s\hat{\sigma}_j^+ - \hat{\sigma}_i^+\hat{\sigma}_j^-\hat{\rho}_s - \hat{\rho}_s\hat{\sigma}_i^+\hat{\sigma}_j^-).
\end{equation}
When $\Delta_c \gg \kappa/2$, the dynamics is approximately unitary.\\

For realistic state of the art experiments, $g\sim 2 \pi \times 4$ Hz, $\Delta_c\sim 1$Mhz $J_{\rm z}\sim 21$Hz, and $J = g^2/(\Delta_c J_z) \sim 0.76$ \cite{norcia2018cavity,borish2020transverse}; the spin chain exhibits fast scrambling in this parameter regime (see Fig.~\ref{fig_last}). These results indicate that our proposal provides a promising avenue for observing fast scrambling in state-of-the-art quantum simulators.

\bibliography{ref} 
%\bibliography{reference}

\end{document}